\documentclass[twocolumn,amsmath,amssymb,superscriptaddress,nofootinbib,longbibliography,prl]{revtex4-1}

\usepackage{graphicx}
\usepackage{bm}
\usepackage{dsfont}
\usepackage[usenames,dvipsnames]{xcolor}
\usepackage{pstricks}
\usepackage[tight]{subfigure}
\usepackage{verbatim}
\usepackage{units}
\usepackage{natbib}
\usepackage{multirow}
\usepackage{enumitem}
\usepackage{mathrsfs}
\usepackage{leftidx}
\usepackage{xspace}

\usepackage[normalem]{ulem}  

\usepackage[a4paper]{hyperref}
\hypersetup{colorlinks=true,linktoc=all,linkcolor=blue,breaklinks=true,citecolor=blue,urlcolor=blue}

\usepackage[english]{babel}
\AtBeginDocument{%
    \newwrite\bibnotes
    \def\bibnotesext{Notes.bib}
    \immediate\openout\bibnotes=\jobname\bibnotesext
    \immediate\write\bibnotes{@CONTROL{REVTEX41Control}}
    \immediate\write\bibnotes{@CONTROL{%
    apsrev41Control,author="08",editor="1",pages="1",title="0",year="1"}}
     \if@filesw
     \immediate\write\@auxout{\string\citation{apsrev41Control}}%
    \fi
}%

\definecolor{DarkGreen}{rgb}{0,0.4,0}

\usepackage{hyperref}
\begin{document}
	
	
	\title{Non-equilibrium System as a Demon}
	
	\author{Rafael S\'anchez}
	\affiliation{Departamento de F\'isica Te\'orica de la Materia Condensada and Condensed Matter Physics Center (IFIMAC), Universidad Aut\'onoma de Madrid, 28049 Madrid, Spain}

	\author{Janine Splettstoesser}
	\affiliation{Department of Microtechnology and Nanoscience (MC2), Chalmers University of Technology, S-412 96 G\"oteborg, Sweden\looseness=-1}

	\author{Robert S. Whitney}
	\affiliation{Laboratoire de Physique et Mod\'elisation des Milieux Condens\'es, Universit\'e Grenoble Alpes and CNRS, BP 166, 38042 Grenoble, France}

	\date{\today}                                                                                                                                                                                                                                                 
	\begin{abstract}
Maxwell demons are creatures that are imagined to be able to reduce the entropy of a system without performing any work on it. Conventionally, such a Maxwell demon's intricate action consists of measuring individual particles and subsequently performing feedback. Here we show that much simpler setups  
can still act as demons: we demonstrate that it is sufficient to exploit a non-equilibrium distribution to \emph{seemingly} break the second law of thermodynamics. We propose both an electronic and an optical implementation of this phenomenon, realizable with current technology.
	\end{abstract}

	\maketitle	


\textit{Introduction.} 
The second law of thermodynamics requires entropy to increase on long time scales. Maxwell demons apparently break this law by decreasing the entropy in a system without transferring any energy to it~\cite{Maxwell}. 
They do this by measuring individual particles and performing feedback based on the information acquired. 
The second law is restored \cite{Bennett1982Dec} by the Maxwell demon generating entropy when erasing the information it has acquired about the system~\cite{Landauer1961Jul}.
Even though highly intricate, such Maxwell  demons have been built in electronic~\cite{Koski2014Jul,Koski2014Sep,Koski2015Dec,Chida2017May}, superconducting~\cite{Cottet2017Jul,Masuyama2018Mar}, and optical~\cite{Vidrighin2016Feb} systems, using NMR~\cite{Camati2016Dec}, and optically or electrically controlled molecules~\cite{Serreli2007Feb} or microscopic objects~\cite{Toyabe2010Nov,Berut2012Mar,Roldan2014Apr}. 

\begin{figure}[b]
\includegraphics[width=0.9\columnwidth]{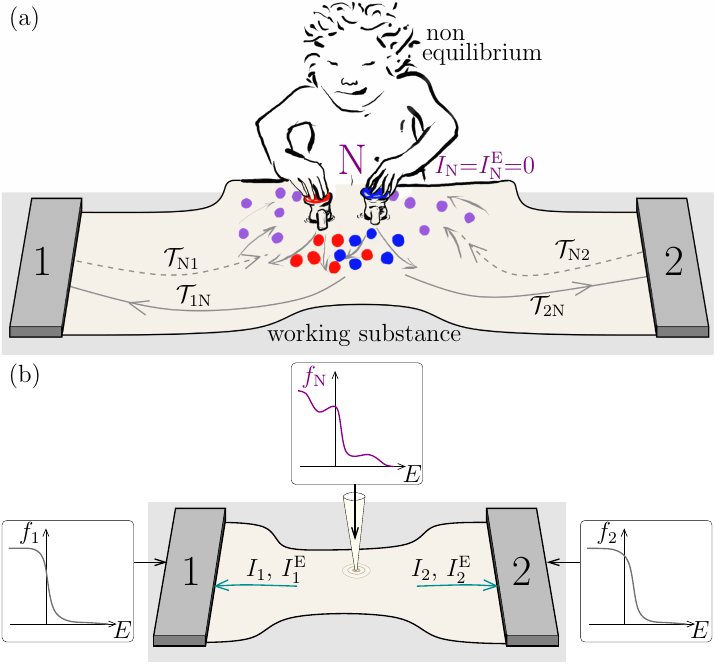}
\caption{\label{fig_setup}
(a) The N-demon supplies no heat or work, but a non-equilibrium distribution to the working substance containing equilibrium reservoirs 1 and 2.  The non-equilibrium distribution could be a non-thermalized mixture of different equilibrium distributions.
Transmission probabilities, $\mathcal{T}_{ij}$, of the scattering region (beige) involving terminal N are indicated and accompanied by gray arrows. (b) A physical implementation in which the non-equilibrium distribution $f_{\rm N}$ is injected locally into the working substance.
}
\end{figure}

This Letter shows that a much simpler class of setups has an analogous effect, \textit{without} 
involving any measurement of individual particles (namely avoiding any acquisition of information), or any feedback, but instead exploiting a non-equilibrium (N) distribution.  
In this sense such a setup is different from a typical Maxwell demon, and we call it an N-demon.
This N-demon induces a {\it steady-state} reduction of the expectation value of the entropy of a pair of reservoirs, $\dot{S}_1+\dot{S}_2 <0$,
without any steady-state supply of heat, work, or other energy.  
We consider two examples of this entropy reduction:
\begin{itemize}[noitemsep,topsep=0pt]
\item[(i)] Heat in reservoirs 1 and 2 is turned into work when the two reservoirs are at the same temperature, generating electrical (or electrochemical) power while cooling these reservoirs.
\item[(ii)] Heat is moved from reservoir 1 to 2, when reservoir~1 is colder.
\end{itemize}
These are, respectively, apparent violations of the Kelvin and Clausius versions of the second law.
An example of such a N-demon is shown in Fig.~\ref{fig_setup}(a). 
This demon injects a non-equilibrium distribution of particles symbolized by a pair of faucets inserting particles from cold (blue) and  hot (red) distributions, at rates chosen so that they carry the same average energy as the backflow of particles (magenta). This way, there is no steady-state particle or  energy  flow between the N-demon and the working substance.
In other words, there is no flow of heat or work. 
We propose straightforward implementations of such N-demons in both electronic and optical systems, and show that they do not violate the second law.

 \textit{General entropic analysis.} 
Consider the three-terminal set-up in Fig.~\ref{fig_setup}, where our aim is that
terminal N~\cite{terminology-terminal} 
(the N-demon)
reduces the entropy of the working substance  (always indicated by a gray background) containing reservoirs 1 and 2. 
We are interested in exploiting terminal N's non-equilibrium
distribution as the resource, unlike traditional thermodynamics which exploit heat as the resource.  To clarify the effect of the non-equilibrium distribution alone, we will concentrate on cases where terminal N supplies no heat or work to reservoirs 1 and 2; we will call this the ``demon conditions'' below.

We assume reservoirs 1 and 2 are each in internal equilibrium, so the rate of entropy change in each is given by a Clausius relation $\dot{S}_i=J_i/T_i$, where $J_i$ is the heat current into reservoir $i{=}1,2$, which has temperature $T_i$.  
For particles in the presence of an electrochemical potential $\mu_i$, the heat current is $J_i=I^\text{E}_i{-}\mu_iI_i$ with the particle current $I_i$ and the energy current $I^\text{E}_i$~\cite{Benenti2017Jun,Whitney2018May}.
However, as N is out of equilibrium, there is no such relationship for $\dot{S}_\text{N}$. 
The second law of thermodynamics 
is 
\begin{equation}\label{eq:entropy_production}
0 \,\leq\, \dot{S}_\text{N} \,+\, J_1 \big/ T_1 \,+\, J_2\big/T_2 \ .
\end{equation}
The ``demon conditions'' that the N-demon 
neither injects or extracts heat or work, are   $I^\text{E}_\text{N}=I_\text{N}=0$.
If the N-demon were in internal equilibrium, it too would obey a Clausius relation, so these conditions would fix $\dot{S}_\text{N}=0$.
Then Eq.~(\ref{eq:entropy_production}) would become the usual 
second law for two reservoirs, forbidding the reduction of the sum of their entropies.
However, 
one can have $\dot{S}_\text{N}\neq0$ under demon conditions, if terminal N is out of equilibrium.

Take example (i) above, with reservoirs 1 and 2 at the same temperature $T$, but with $\mu_1\neq \mu_2$. 
The second law in Eq.~(\ref{eq:entropy_production}) becomes
$P = (\mu_1-\mu_2)I_1 \leq T \dot{S}_\text{N}$,
where $P$ is the electrical power output.   Thus if $\dot{S}_\text{N}$ is positive, the working substance is allowed to do work (positive $P$) even when the N-demon supplies no work or heat, 
$I^\text{(E)}_\text{N}=I_\text{N}=0$. This means the work output comes from a reduction of heat in reservoirs 1 and 2,  $J_1+J_2= -P<0$, in apparent violation of Kelvin's second law.

For example (ii) above, $T_1\neq T_2$ but $\mu_1=\mu_2$. Then 
Eq.~(\ref{eq:entropy_production}) becomes $J_1 (T_1-T_2) \leq T_1 T_2 \dot{S}_\text{N}$ under demon conditions.
So when $\dot{S}_\text{N}$ is positive, heat may flow from cold to hot (i.e. $J_1$ may have the opposite sign to $T_2-T_1$), even though no energy comes from the N-demon, in apparent violation of Clausius' second law.

These arguments show that the demon effects (i) and (ii) do not violate the laws of thermodynamics. 
Note that to fix the demon conditions one requires knowledge of the steady-state flows of charge and energy, $I^\text{E}_i$ and $I_i$, but not of the behavior of individual particles. In particular, the N-demon operates without needing to know about any microscopic details of the working substance.
Once the demon conditions are fixed, the N-demon generates work indefinitely, without any further measurement or adjustment. 

In the rest of this Letter we propose two systems which indeed exhibit such effects.
Crucially, throughout this Letter, we assume noninteracting particles, which excludes any interpretation in terms of autonomous feedback~\cite{Esposito2012Aug,Strasberg2013Jan,Barato2013Mar,Shiraishi2015Apr,Koski2015Dec}.
This is the critical difference from a similar setup with strong Coulomb interactions \cite{Whitney2016Jan}, 
which can be understood as an autonomous Maxwell demon \cite{Hofer-Sanchez}.

\begin{figure}[t]
\includegraphics[width=3.in]{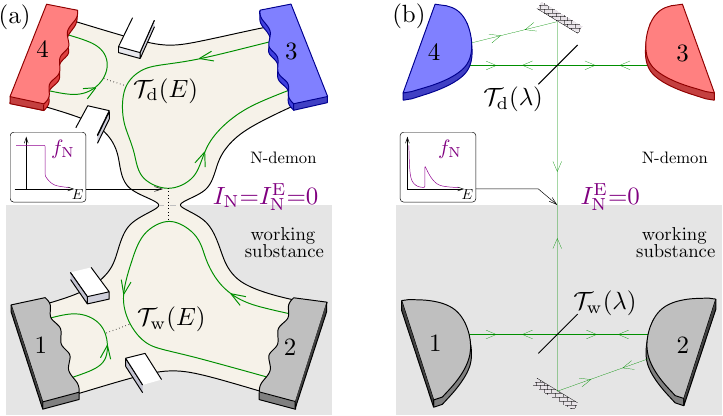}
\caption{\label{fig_concrete}
(a) Electronic, quantum-Hall bar (beige) with two constrictions  with energy-dependent transmissions $\mathcal{T}_\text{d}(E)$ of the N-demon and $\mathcal{T}_\text{w}(E)$ of the working substance. The four reservoirs have different temperatures $T_i$ and electrochemical potentials $\mu_i$. (b) Optical setup with four black-bodies with temperatures $T_i$ and wavelength-dependent half-silvered mirrors with $\mathcal{T}_\text{d}(\lambda)$ and $\mathcal{T}_\text{w}(\lambda)$. 
}
\end{figure}

\textit{Scattering description.} 
As there are no interparticle interactions, the setups of interest can be described using scattering theory~\cite{Butcher1990,Buttiker1992Nov}, which is known to respect the second law~\cite{Nenciu2007Mar,Whitney2013Mar,Benenti2017Jun}.
The particle and energy currents into reservoir $i$ are $I_i{=}I^{(0)}_i$ and $I^\text{E}_i{=}I^{(1)}_i$, where
\begin{equation}\label{eq:scattering_currents}
I^{(\nu)}_i=\frac{1}{h}\sum_{j}\sum_{k,k'}\int dE \ E^\nu \ \mathcal{T}^{kk'}_{ij}(E) \,\left[f_j(E){-}f_i(E)\right].
\end{equation}
 Here, $\mathcal{T}^{kk'}_{ij}(E)$ are the transmission probabilities from channel $k'$ in $j$ to channel $k$ in  $i$ at energy $E$, see Fig.~\ref{fig_setup} (superscripts $k,k'$ are dropped when not relevant). 
For equilibrium reservoirs, $f_i(E)$ are Fermi or Bose distributions, depending on the discussed setup.
Importantly, the demon effect requires that the non-equilibrium terminal N has asymmetric and energy-dependent 
couplings to reservoirs 1 and 2, with  $\mathcal{T}_{1\text{N}}(E)\neq \mathcal{T}_{2\text{N}}(E)$ for at least some energy $E$.  
The electronic and optical setups proposed in Fig.~\ref{fig_concrete} fulfill these requirements. 

For simplicity, in these setups the nonequilibrium distribution is created from mixing the flows from two equilibrium reservoirs. Note however, that no spatial separation of these two flows is required. Hence, for clarity, the proposed setups have the demon and the working substance exchanging particles at a single point.

\textit{Proposed implementation 1: Quantum-Hall setup.} 
As an electronic implementation, we propose a quantum-Hall bar in contact with four reservoirs, in which electron transport takes place via chiral edge states~\cite{Buttiker1988Nov}, marked by green lines with arrows in Fig.~\ref{fig_concrete}(a). Such a setup is in experimental reach, since effects of non-equilibrium distributions~\cite{Altimiras2009Oct} and heat current measurements~\cite{Jezouin2013Nov} have been demonstrated.
We focus on the N-demon (reservoirs 3 and 4 \textit{together} taking the role of the non-equilibrium terminal N in Fig.~\ref{fig_setup}). The demon is connected by a constriction (with energy-independent transmission, taken for simplicity to be equal to one) to the working substance, which generates work as in example (i) above. 
The work is electrical, with the N-demon moving electrons against the potential difference $\mu_2{-}\mu_1$ between reservoirs 1 and 2.
The non-equilibrium distribution which performs this demonic action is formed using the equilibrium distributions from reservoirs 3 and 4 with possibly different temperatures
$T_{3/4}=T{+}\delta T_{3/4}$ 
and electrochemical potentials, $\mu_{3/4}=\mu{+}\delta\mu_{3/4}$.
Mixing by an energy-dependent scatterer with transmission $\mathcal{T}_\text{d}(E)$, yields $f_\text{N}(E)=\left[1{-}\mathcal{T}_\text{d}(E)\right]f_3(E)+\mathcal{T}_\text{d}(E)f_4(E)$. 
Two of the four parameters ($\delta T_3$, $\delta T_4$, $\delta\mu_3$ and $\delta\mu_4$) determine the out of equilibrium distribution, while the other two are tuned to ensure the demon conditions, $I_3{+}I_4:=I_\text{N}=0$ and $I_3^\text{E}{+}I_4^\text{E}:=I^E_\text{N}=0$.
These demon conditions are similar to the condition for a voltage or temperature probe~\cite{Buttiker1986Oct,Buttiker1988Nov,Buttiker1988Jan}, but we repeat that they should not be confused with the measurement-feedback scheme of a standard Maxwell demon. 

The energy-dependent transmission asymmetry is done by inserting a 
scatterer with transmission $\mathcal{T}_\text{w}(E)$, leading to transmission probabilities $\mathcal{T}_{14}(E){=}\mathcal{T}_\text{d}(E)\mathcal{T}_\text{w}(E)$ and $\mathcal{T}_{21}(E){=}\mathcal{T}_\text{w}(E)$. 
In the linear regime (small potential and temperature differences), analytic results are as follows. 
For affinities $F_i^\mu=\delta\mu_i/k_\text{B}T$, and $F_i^T=
\delta T_i/k_\text{B}T^2$, and  defining $F_{ij}^x=F_i^x-F_j^x$,
the demon conditions imply that
\begin{eqnarray}\label{eq:V_linear}
F_{32}^\mu&=&F_3^T\frac{(g_\text{d}^{1})^2+g_\text{d}^{0}X_{0\text{d}}^{2}}{g_0^{0}g_\text{d}^{1}}
-F_4^T\frac{(g_\text{d}^{1})^2-g_\text{d}^{0}g_\text{d}^{2}}{g_0^{0}g_\text{d}^{1}}
\end{eqnarray}
and that $F_{42}^\mu$ takes the same form with $g_\text{d}^{0}$ replaced by $-X_{0\text{d}}^{0}$.
Here $X_{\alpha\beta}^{\nu}=g_\alpha^{\nu}-g_\beta^{\nu}$  for $\alpha,\beta=\text{d,w},0$, with 
$g_{\alpha...\beta}^{\nu}=(k_\text{B} T/h)\int dE \left({-}\partial_E f\right)E^\nu\mathcal{T}_\alpha{\ldots}\mathcal{T}_\beta$ for Fermi function $f$ and $\mathcal{T}_0=1$ . 
Then the particle currents are 
\begin{eqnarray}\label{eq:current_linear}
I_1=-I_2=g_\text{w}^0F_{21}^\mu+F^T_3\frac{g_0^2(g_\text{w}^0g_\text{d}^0-g_\text{dw}^0g_0^0)+g_0^0g_\text{w}^1g_\text{d}^1}{g_0^0g_\text{d}^1}\nonumber\\
-F^T_{43}{\left[g_\text{w}^0\frac{(g_\text{d}^1)^2-g_\text{d}^2g_\text{d}^0}{g_\text{d}^1g^0_0}-\frac{g_\text{d}^1g_\text{dw}^{1}-g_\text{dw}^0g_\text{d}^2}{g_\text{d}^1}\right]}{.}\ 
\end{eqnarray}
The crucial point is that the second and third term on the right hand side can overcome the first term, such that a current flows between reservoirs 1 and 2 against the potential gradient. 
Algebra shows that this only occurs for transmissions with 
the properties below Eq.~(\ref{eq:scattering_currents}).

In stark contrast with known thermoelectric generators~\cite{Benenti2017Jun,Whitney2018May,Sanchez2015Apr}, this works even if the working substance is electron-hole symmetric 
($\mathcal{T}_\text{w}(E)$ symmetric about $\mu$).  
The non-equilibrium distributions violate the electron-hole symmetry (via $\mathcal{T}_\text{d}(E)$), 
so one can have finite power output even when $\mathcal{T}_\text{w}(E)$ is symmetric.

\begin{figure}[tb]
\includegraphics[width=3.3in]{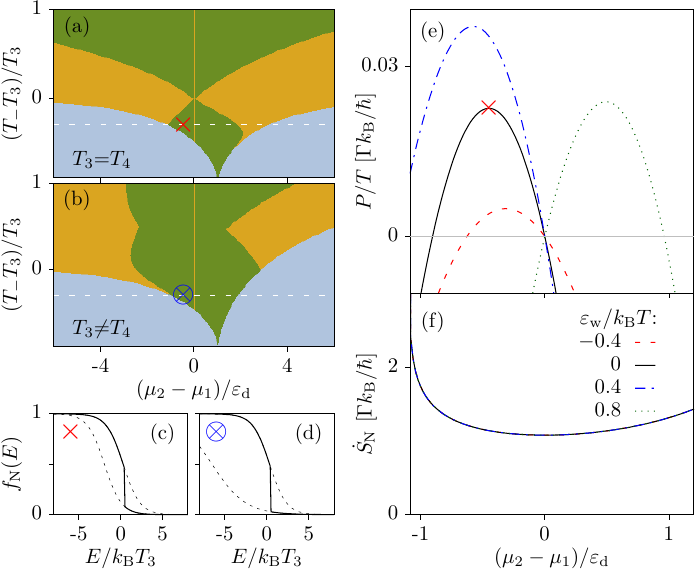}
\caption{\label{fig_QH}
Power generation in the working substance of the quantum-Hall setup. $\mathcal{T}_{\rm d}$ is a step function with threshold $\varepsilon_\text{d}/k_\text{B}T_3{=}0.5$, and $\mathcal{T}_{\rm w}$ has a  resonance $\varepsilon_{\rm w}$ with width $\Gamma$. (a) $T_3{=}T_4$: Map of regions where the demon conditions cannot be fulfilled (cyan), can be fulfilled, but no power is generated (yellow), and where power generation with the N-demon is possible (green). At \textcolor{red}{$\times$} the N-demon produces maximum power if $\varepsilon_\text{w}{=}0$; the injected non-equilibrium distribution is shown in (c) compared to $f_{3/4}(E)$ (dashed). (b) Same as (a) with $T_4{=}2T_3$. The  injected non-equilibrium distribution at \textcolor{blue}{$\otimes$} is shown in (d). (e) Power generated in the working substance and (f) entropy  production in the N-demon, for $T_3{=}T_4=7\ T/10$ (white dashed line in (a)) and different  $\varepsilon_\text{w}$.
}
\end{figure}

Fig.~\ref{fig_QH} presents results for the nonlinear regime, from Eq.~\eqref{eq:scattering_currents} for arbitrary $T_i$ and $\mu_i$. 
We choose the N-demon's transmission as a quantum point contact, $\mathcal{T}_\text{d}(E)=\theta(E-\varepsilon_\text{d})$,
and the working substance's transmission to be that of a weakly coupled quantum dot, $\mathcal{T}_\text{w}(E)=\Gamma^2/[(E-\varepsilon_\text{w})^2+\Gamma^2]$ with small width $\Gamma$. 
These are experimentally well-understood and controllable circuit elements. 
For fixed $\mathcal{T}_\text{d}$ and temperatures $T_3$ and $T_4$, Figs.~\ref{fig_QH}~(a) and (b) show the regions where the demon conditions can be met by adjusting $\mu_3$ and $\mu_4$ and where power generation is possible 
with $\varepsilon_\text{w}$ tuned to a suitable value with gates. The shape of these maps depends on $\mathcal{T}_\text{d}$, however they usually show power generation under demon conditions in an extended parameter regime, even when  $T_3=T_4$, so long as $T_3\neq T$, see  Fig.~\ref{fig_QH}(a).

Taking $T_3=T_4=7\ T/10$, Fig.~\ref{fig_QH}~(e) shows the power generated as a function of $\mu_1-\mu_2$ and for different $\varepsilon_\text{w}$. For two specific situations, we show the non-equilibrium distribution injected by the N-demon, see Figs.~\ref{fig_QH}(c) and (d).
 As required by Eq.~\eqref{eq:entropy_production},  the entropy production in the N-demon, $\dot{S}_\text{N}=J_3/T_3+J_4/T_4$, is always larger than $P/T$, see Fig.~\ref{fig_QH}(f).  However, in contrast to what one would expect from a feedback-based demon~\cite{Esposito2012Aug,Strasberg2013Jan,Barato2013Mar,Shiraishi2015Apr,Koski2015Dec}, the entropy production of the N-demon does not depend on the details of the working substance.
 Importantly, this entropy production is spatially completely separated from the working substance  and its control (or even minimization) is hence of minor relevance.

\begin{figure}[bt]
\includegraphics[width=3.3in]{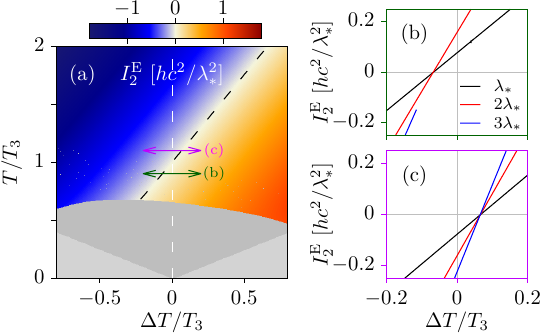}
\caption{\label{fig_optical}
Optical N-demon. (a) Cooling power (energy current into reservoir 2) as function of $\Delta T$ and $T$, for $T_{1(2)}=T\pm\Delta T$, fixed $T_3$.
We set $\lambda_0=\lambda_*$, and $\lambda_{ b}=0.02\lambda_*$, $\lambda_{a}=5\lambda_*$.  $T_4$ is given by the demon conditions, which cannot be fulfilled in the dark-gray region (light gray regions have unphysical negative $T_{1/2}$). White/black dashed lines at $\Delta T=0$ and at vanishing cooling power are shown as guides for the eye. Black lines in  (b) and (c) show cuts marked by green and magenta arrows at $T=0.9T_3$ and $T=1.1T_3$, respectively,  together with results for different $\lambda_0=2\lambda_*,3\lambda_*$ but the same $T$ (lines stop when demon conditions are unfulfillable). 
}
\end{figure}

\textit{Proposed implementation 2: Optical setup.} 
Fig.~\ref{fig_concrete}(b) shows a N-demon implementation of example (ii) above in an optical setup with non-interacting photons~
\cite{photons}.
The demon-part of the setup consists of two thermal (black-body) photon sources at temperatures $T_3$ and $T_4$, emitting light in a wavelength window $[\lambda_b,\lambda_a]$ (respectively an energy window $[E_a,E_b]$ with $E_{a,b}=hc/\lambda_{a,b}$).
Both emit photons onto a mirror, which transmits or reflects light in a wavelength-selective manner $\mathcal{T}_\text{d}(\lambda)$.  
The resulting non-equilibrium distribution is sent into the lower part of the device, the working substance. The latter consists of two black-bodies with a temperature difference $\Delta T=T_1-T_2$.  The relation between temperatures required to satisfy the demon condition, $I^\text{E}_3+I^\text{E}_4=0$, depends on the N-demon's transmission $\mathcal{T}_\text{d}(\lambda)$ \textit{and} the transmission $\mathcal{T}_\text{w}(\lambda)$ of the working substance. 
In linear response, the demon conditions reduce to 
$F_3^T =  \left[\left(g_0^2-g^2_\text{w}\right)F_1^T+g_\text{w}^2F_2^T-g_\text{d}^2F_4^T\right]\big/
\left(g^2_0-g^2_\text{d}\right)$
with the same abbreviations as for the electronic setup, but $f_i(E)\equiv f_i(hc/\lambda)$ being Bose distributions. 
Then
\begin{eqnarray}
I_2^\text{E} & = & \left[A F_{12}^T  - g^2_0g_\text{dw}^2 F_{14}^T + g_\text{d}^2g_\text{w}^2 F_{24}^T\right]\Big/ \left(g^2_0-g^2_\text{d}\right) 
\label{eq:heat_linear}
\end{eqnarray}
where $A=g_\text{w}^2 \left(2g_0^2-g_\text{d}^2 -g_\text{w}^2+g_\text{dw}^2\right)$.
A simple example shows that heat flow between reservoirs 1 and 2 is not always from hotter to colder. Fixing the wavelength-dependent transmissions to be $\mathcal{T}_\text{d}(\lambda)=\theta(\lambda-\lambda_0)=1-\mathcal{T}_\text{w}(\lambda)$
 we have $I^\text{E}_2\rightarrow g_0^1F_{12}^T+g_\text{d}^1F_{24}^T$. Then heat flows from cold to hot when $F_{12}^T$ and $F_{24}^T$ have opposite signs and the magnitude of $F_{24}^T$ compensates for the difference between $g_0^2$ and $g_\text{d}^2$.
 Fig.~\ref{fig_optical}(a) shows the full, nonlinear energy current into reservoir 2 as function of $T$ and $\Delta T$. 
 Cooling of the colder reservoir occurs between the dashed lines.
 Figure panels~\ref{fig_optical}(b) and (c) show line plots of the cooling power (black lines) for two
 examples at fixed temperatures $T$ (indicated by arrows).  
They also show that the cooling power is enhanced by tuning $\lambda_0$.
 We have assumed that each frequency contributes with a single spatial mode. An increase of the overall cooling power is expected when increasing the mode number.
 
\textit{Requirements for experimental demonstrations.} For the quantum-Hall implementation of the N-demon, we expect power outputs of the order of $P\approx10$ aW, when choosing $T=70$mK, $T_3=T_4=100$mK, and $\Gamma=1\mu e$V $(\approx 0.1k_\text{B}T)$. In the optical setup in the near-infrared regime, with wavelengths $\lambda_\text{d}=\lambda_\text{w}=6.2\mu$m, $\lambda_a=12.4\mu$m, and $\lambda_b=0.25\mu$m and temperatures around $T=1000$K, the cooling power changes between 0 and $\pm0.1\mu$W for temperature gradients between $\pm100$K and 0, respectively. These numbers are experimentally attainable. It is also necessary to experimentally demonstrate that the device is operating under demon condition. For the electronic setup in Fig.~\ref{fig_setup}~(a), a quantum dot with a delta-shaped transmission $\mathcal{T}_\text{w}(E)$, was chosen for the example studied in Fig.~\ref{fig_QH}, because it allows a read out~\cite{Altimiras2009Oct} of the incoming non-equilibrium distribution function. 
An additional side-coupled dot could be used at the outgoing channel from reservoir 2 to 3 to monitor the re-injected equilibrium distribution.
In the optical setup, the incoming and outgoing light from the N-demon can be split by a mirror and sent on separate spectrum analyzers. From the detected distribution functions  particle and energy currents can be deduced. 

\textit{Practical uses.}  
Our demons are less intricate to construct than standard Maxwell demons, so their practical uses merit consideration.  The implementations that we suggest could be used to spatially separate\emph{production} from \emph{reduction} of entropy, for nanoscale heat management  This is a more general version of the non-locality of thermodynamics laws identified in Ref.~\cite{Whitney2016Jan}.  
Crucially, if some other independent process generates a non-equilibrium distribution as "waste", our results show that one can use its
non-equilibrium nature as a resource to perform work (or cooling). 

\textit{Conclusion.} We have shown that non-nequilibrium distributions can be exploited for power generation and cooling 
in a demon-like manner. In contrast to other demon-like devices exploiting ``engineered reservoirs'', see e.g.\  Refs.~\cite{Scully2003Feb,Francica2017Mar,Ghosh2018Mar}, our proposal does not require any subtle quantum coherence or correlation effects.
We have proposed two very different implementations,
that could be constructed with current technology.   
For clarity, our two examples have their non-equilibrium distributions made out of two equilibrium reservoirs,
however nature is rife with other types of non-equilibrium systems. 
Our thermodynamics arguments imply that generic non-equilibrium systems could act as N-demons.
It is sufficient that the demon and the working substance exchange
energy, similar to Ref.~\cite{Whitney2016Jan}; there is no requirement for the particle exchange.
One could also have hybrid systems, e.g.\ an optical N-demon acting on an electronic working substance. 
Transient non-equilibrium effects may also be of interest~\cite{Konopik2018Jun}.   

\acknowledgements
\textit{Acknowledgements.}
We acknowledge stimulating discussions with the participants of the KITP program ``Thermodynamics of Quantum Systems". This research was supported in part by the National Science Foundation under Grant No. NSF PHY-1748958 (RS and JS), by the Knut and Alice Wallenberg Foundation and the Swedish Vetenskapsradet VR (JS), by the Spanish MINECO via grant FIS2015-74472-JIN (AEI/FEDER/UE), the MAT2016-82015-REDT  network and the program MDM-2014-0377, and the Ram\'on y Cajal program RYC-2016-20778 (RS), and the French ANR-15-IDEX-02 via the Universit\'e Grenoble Alpes QuEnG project (RW).


\begin{thebibliography}{39}
\bibitem [{ter()}]{terminology-terminal}%
  \BibitemOpen
  \href@noop {} {}\bibinfo {note} {We use the word ``terminal'' for N, because
  ``reservoir'' often implies internal equilibrium, which is not appropriate for
  terminal N.}\BibitemShut {Stop}%
\bibitem [{pho()}]{photons}%
  \BibitemOpen
  \href@noop {} {}\bibinfo {note} {The constraint of zero particle flow is
  dropped here, because photons are not conserved in the reservoirs, so they
  have zero chemical potential. This means that particle flow cannot affect
  the thermodynamics of heat and work.}\BibitemShut {Stop}%
\end{thebibliography}

%

\end{document}